\documentclass[12pt,preprintnumbers,aps,amssymb,nofootinbib,amsmath]{revtex4}
\usepackage{epsfig,epsf}
\usepackage{graphicx}
\usepackage{epstopdf}
\newcommand{\beq}{\begin{equation}}
\newcommand{\eeq}{\end{equation}}
\def\eq#1{{(\ref{#1})}}
\def\fig#1{{Fig.~\ref{#1}}}

\newcommand{\as}{\alpha_s}

\newcommand{\lag}{\mathcal{L}}

\newcommand{\bea}{\begin{eqnarray}}
\newcommand{\eea}{\end{eqnarray}}
\newcommand{\ben}{\begin{eqnarray*}}
\newcommand{\een}{\end{eqnarray*}}
\newcommand{\vac}{|\epsilon_\mathrm{v}|}

\begin{document} 

\preprint{BNL-NT}
\preprint{RBRC-757}
\preprint{TAUP-2882-08}

\title{Broken scale invariance, massless dilaton and confinement in QCD}
%\title{An effective theory of broken scale invariance in QCD at intermediate distances: 
%massless dilaton  and confinement}

\author{Dmitri Kharzeev$^a$, Eugene Levin$^b$ and Kirill Tuchin$^{c,d}$}
 
\affiliation{
a) Department of Physics, Brookhaven National Laboratory,\\
Upton, New York 11973-5000, USA\\
b) HEP Department, School of Physics and Astronomy,\\
Tel Aviv University, Tel Aviv 69978, Israel\\
c) Department of Physics and Astronomy,\\
Iowa State University, Ames, Iowa, 50011, USA\\
d) RIKEN BNL Research Center,\\
Upton, New York 11973-5000, USA\\
}

\date{\today}

\begin{abstract} 
Classical conformal invariance of QCD in the chiral limit is broken explicitly by scale anomaly. 
As a result, the lightest scalar 
particle (scalar glueball, or dilaton) in QCD is not light, and cannot be described as a Goldstone boson. Nevertheless basing on an effective low-energy theory of broken scale invariance we argue that inside the hadrons the non-perturbative interactions of gluon fields result in the emergence of a massless dilaton excitation (which we call the ``scalaron"). We demonstrate that our effective theory of broken scale invariance leads to confinement. This theory allows a dual formulation as a classical Yang-Mills theory on a curved conformal space-time background. Possible applications are discussed, including the description of strongly coupled quark-gluon plasma and the spin structure of hadrons.  
\end{abstract}

\maketitle 
\newpage

%%%%%%%%%%%%%%%%%%%%%%%%%%%%%%%%%%%%%%%%

\section{Introduction and Summary}

The relevant degrees of freedom in QCD at large distances are still poorly understood. 
A perturbative approach in this domain is rendered impossible by the lack of convergence and by the existence of multiple solutions to the gauge fixing condition (Gribov copies) \cite{Gribov:1977wm}. 
Quasi-classical solutions are also poorly defined at strong coupling, and may not  dominate the amplitudes in the presence of large quantum corrections. 

A possible way of making QCD treatable at large distances is to use an effective theory encoding the known symmetries of the Lagrangian. At very large distances, the effective degrees of freedom are pions  and the corresponding effective theory is based on spontaneously broken chiral invariance. Another symmetry present in QCD in the chiral limit  
is  scale invariance -- indeed, in the limit of massless quarks the Lagrangian does not possess 
any dimensionful parameters and is therefore invariant with respect to rescaling of coordinates, 
$x_{\mu} \to \lambda x_{\mu}$. In real world, this invariance is lost -- the hadrons possess finite masses and sizes, so the scale symmetry is broken. It is therefore tempting to formulate an effective theory of broken scale invariance in terms of the corresponding Goldstone boson - the dilaton. Since the scale transformation does not affect any quantum numbers, the corresponding particle has to be a scalar -- a scalar glueball or perhaps, in the presence of light quarks, a $\sigma$ or $f_0$ meson. 

There is a well-known problem with this approach however: unlike chiral symmetry, the scale invariance is broken not spontaneously, but explicitly \cite{scale1,scale2}. Indeed the quantum effects and the regularization required to treat them bring in a dimensionful constant -- $\Lambda_{\rm QCD}$ \cite{Gross:1973id}. As a result, there is no limit in which the dilaton can become massless. Nevertheless, the corresponding effective theory of a scalar field 
can still be formulated \cite{schechter,Migdal:1982jp}-- it is fixed unambiguously by the Ward identities of broken scale invariance. 
However how useful this theory can be depends on how softly the scale symmetry is broken, and how 
massive the resulting dilaton is. In gluodynamics, the dilaton is identified with the scalar glueball. The mass of the scalar glueball according to the lattice calculations is quite large, $M_G \simeq 1.5 \div 1.7$ GeV \cite{Chen:2005mg}. While this {\it is} the lightest glueball, its mass is not significantly smaller than the masses of glueballs with other quantum numbers \cite{Chen:2005mg}, so there is no reason to expect that the physics at large distances will be dominated by the dynamics of scalar glueballs.  Moreover, the corresponding Compton wavelength $\sim M_G^{-1}$ is much shorter than the confinement radius $\sim \Lambda_{\rm QCD}^{-1}$, so 
it is likely that the dynamics described by the effective theory of glueballs is entangled with the gluon dynamics. 
This calls for an extension of the effective theory to include both scalar fields and gluons, and such an extension has been carried out in Refs.\cite{klt,Kharzeev:2004ct}.   
The resulting theory involves the scalar dilaton field $\chi$ interacting with the gluon field characterized by $F_{\mu\nu}^a$. 

\medskip

In this paper we will explore the dynamics of this model further; let us briefly summarize our findings. 
In the absence of color fields, the minimum of the effective potential for the scalar field $\chi$ is at $\chi = 0$, and the 
Lagrangian describes the theory of self-interacting scalar glueballs. However as the strength 
of the color field increases, the minimum at $\chi = 0$ disappears and 
a non-zero expectation value for $(F_{\mu\nu}^a)^2 < 0$ develops, corresponding to the formation of a  confining chromo-electric flux tube. At the same time the potential for the field $\chi$ vanishes, and the dilaton excitation becomes massless. 

While the resulting theory of gauge bosons interacting with a scalar field may look similar to a 
Higgs model, there is a big difference -- the presence of dilatons does not break the gauge symmetry. Instead, the propagator of gluons in Coulomb gauge acquires a infrared divergent piece $\sim 1/k^4$ 
that corresponds to linear confinement in coordinate space. Such a propagator has been shown 
to emerge once the multiple solutions of gauge fixing condition (Gribov copies) are eliminated \cite{Gribov:1977wm}; a confining Coulomb propagator was shown to be a necessary condition for confinement \cite{Zwanziger:2002sh} (for a review, see e.g. \cite{Dokshitzer:2004ie}). The presence of confinement in the model  of \cite{klt,Kharzeev:2004ct} has been previously  investigated by a different method in Ref. \cite{Gaete:2007zn}.

The effective theory of \cite{klt,Kharzeev:2004ct} allows a dual formulation as a classical Yang--Mills theory on a curved conformal space-time background.  Qualitatively, a geometrical interpretation of confinement has been considered in Refs  \cite{Kharzeev:2005iz,Castorina:2007eb}.
\medskip

There are many similarities between this and the previously proposed approaches: the MIT bag model \cite{Chodos:1974je},
the non-topological soliton model of Friedberg and Lee \cite{Friedberg:1976eg},  the Kogut--Susskind model \cite{Kogut:1974sn}, the color dielectric model \cite{Pirner:1991im}, the gauge theory with a dilaton formulated in \cite{Dick:1997ju}, and most of all with the ``perturbative confinement" program of 't Hooft \cite{'t Hooft:2002wz}. Indeed, as we will explain below, the proposed approach  corresponds to the {\it infrared} renormalization of QCD.  Once this 
infrared renormalization is performed, the renormalized gluon propagator possesses the property of confinement. The advantage of the approach presented in this paper is that the structure of the effective theory is completely determined by broken scale invariance of QCD.

In the picture developed in this paper the dominant degree of freedom at intermediate distances 
(distances shorter than the radius of confinement but longer than needed for perturbation theory to apply) 
is the massless scalar excitation composed from gluon fields. We will call it the ``scalaron" to distinguish it from 
the massive glueball, or dilaton state emerging at long distances. The existence of ``scalarons" inside hadrons would have very interesting implications -- for example, since scalarons have zero spin, 
the total spin carried by gluons at intermediate distances inside hadrons should be equal to zero. 
In the physics of dense QCD matter, we expect that close to the deconfinement transition scalarons should play an important dynamical role, possibly inducing a large bulk viscosity in the system in accord with \cite{Kharzeev:2007wb,Karsch:2007jc,Meyer:2007dy}.
 
 \medskip
 
The paper is organized as follows. In section \ref{scaleinv} we introduce the effective Lagrangian of broken scale invariance. In section \ref{sec:conf} we discuss its properties and show that it 
leads to confinement. The structure of the confining gluon propagator is studied in more detail in  
section \ref{sec:el}. In section \ref{sec:thooft} we show that our effective theory can be viewed as a particular realization of 't Hooft's ``perturbative confinement" program. Section \ref{sec:sc} 
contains a discussion of strong coupling behavior in the infrared region, and of the energy density stored in the flux tube. Finally, the discussion of the results is given in section \ref{sec:disc}.

\section{Scale invariance of QCD and the effective theory}\label{scaleinv}

The invariance with respect to the scale transformation $x_\mu \to \lambda x_\mu$ is a property of the QCD Lagrangian in the chiral limit.  Noether's theorem requires the existence of the corresponding conserved dilatation current $s_{\mu}$: $\partial^{\mu} s_{\mu} = 0$. 
Since the divergence of dilatation current in field theory is equal to the trace of the energy-momentum tensor $\partial^{\mu} s_{\mu} = \theta^{\mu}_{\mu}$, in conformally invariant theories $\theta^{\mu}_{\mu} =0$. 
However quantum effects break conformal invariance \cite{scale1,scale2}:
\beq\label{trace0}
\partial^\mu s_\mu=\theta^\mu_{\mu}=\sum_qm_q\,\bar q\,q+\frac{\beta(g)}{2g}
F^{a\mu\nu}F^a_{\mu\nu}\,,
\eeq
where $\beta(g)$ is the QCD $\beta$-function, which governs the behavior of the running coupling
\beq
\mu \frac{d g(\mu)}{d \mu} = \beta (g)\,. \label{rg}
\eeq

At low energy, an effective Lagrangian can be constructed which accounts for  the broken conformal invariance. The simplest possible effective low energy Lagrangian involves one  scalar particle -- \emph{dilaton} \cite{Migdal:1982jp,schechter}. The form of the Lagrangian is uniquely determined by 
the low energy theorems of QCD \cite{Novikov:1981xj}.
   An elegant way to derive this effective Lagrangian has been suggested by Migdal and Shifman in \cite{Migdal:1982jp}. 
They noted that since the gluodynamics is conformally invariant only in four dimensions,  the anomalous contribution to the divergence of the dilatation current appears -- in the dimensional regularization scheme -- as a residual term in the 4D limit. As demonstrated in \cite{Migdal:1982jp}, if we formally couple  the gluodynamics  to the conformal background gravity -- described by a single scalar field $h(x)$ -- such a theory is conformally invariant in any D.  This trick allows to account for all symmetries of the low energy theory. A subsequent Legendre transformation to the conjugate field $\chi(x)$ with a potential that has a classical minimum at $\chi=0$ yields then an effective low energy Lagrangian satisfying all necessary constraints.

This derivation has been generalized in \cite{klt,Kharzeev:2004ct} by including the gluon fields to describe the transition region between the short and long distances. The corresponding Einstein-Hilbert action reads
\beq\label{act}
S\,=\,\int d^4x\,\sqrt{-\mathrm{ g}}\left( \frac{1}{8\,\pi\, G}\, R
 \,-\frac{1}{4} \,
\mathrm{g}^{\mu\nu}\, \mathrm{g}^{\lambda\sigma}\,  F^a_{\mu\lambda}\,
 F^a_{\nu \sigma}\,-\, 
e^{2h}\,\theta_\mu^\mu
\right),
\eeq
where the background metric is given by $\mathrm{g}_{\mu\nu}(x)\,=\,e^{h(x)}\, \delta_{\mu\nu}$, $R$ is the Ricci scalar and $G$ is a dimensionful constant analogous to the Newton's gravitational constant. Upon substitution of the one-loop expression for $\theta_\mu^\mu$ in $SU(3)$ from \eq{trace0} and performing the Legendre transformation we derive the following 
 effective Lagrangian 
\beq\label{LAGR}
\lag \,=\, \frac{\vac}{m^2}\,\frac{1}{2}\,e^{\chi/2}\, (\partial_{\mu}\chi)^2\,
+\,\left(\vac\,+c\,\frac{1}{4} \,(F^a_{\mu\nu})^2  \right) e^\chi\,(1-\chi)\, -\,\frac{1}{4}\,(F^a_{\mu\nu})^2\,;
\eeq
%with $|\chi| \leq 1$; 
the energy density of the vacuum $\vac$ and the mass of the dilaton $m$ are the parameters of the theory.  It is constructed in such a way that at $\chi=1$ (corresponding to some semi-hard momentum scale $M_0$) the terms containing the effective field $\chi$ cancel implying that dynamics of the color fields  is perturbative. This expresses the fact that  $M_0$ is a scale at which the effective theory defined by \eq{LAGR} has to be matched onto the pQCD. The effective theory \eq{LAGR} is non-renormalizable and requires introduction of a cutoff  $M_0(m,\vac)$ at some short distance. 
 In \eq{LAGR} we used notation $c=|\beta(g_0)|/2g_0$  with $g_0\equiv g(M_0)$. Note, that  $c\ll 1$ for any phenomenologically reasonable $g_0$. 
 
 In the absence of color fields, the dilaton potential has a minimum at $\chi =0$ corresponding to 
 the physical vacuum with the energy density $V(\chi =0) = - \vac$. The position of the minimum ($\chi =0$) does not change 
  when the color fields are either predominantly magnetic (corresponding to $(F^a_{\mu\nu})^2 = 2\ (B^{a 2} - E^{a 2}) > 0$) or electric but sufficiently weak so that 
\beq
\vac\,+c\,\frac{1}{4} \,(F^a_{\mu\nu})^2 > 0.
\eeq 
However, the presence of a sufficiently strong color electric field with $(F^a_{\mu\nu})^2 = 2\ (B^{a 2} - E^{a 2}) < 0$ such that $|(F^a_{\mu\nu})^2| > 4 \vac / c\ $ flips the sign of the dilaton potential and the extremum at $\chi=0$ becomes a maximum rather than a minimum. As we will discuss below, this transition corresponds to the formation of color electric flux tube. Since at this point the dilaton potential in \eq{LAGR} vanishes, the formation of the flux tube is accompanied by the emergence of a massless dilaton excitation -- the ``scalaron".
%%%%%%%%%%%%%%%%%%%%%%%%%%%%%%

\section{Confinement by chromo-electric flux tubes}\label{sec:conf}

Consider the Lagrangian \eq{LAGR} which defines our effective low energy 
theory \cite{klt,Kharzeev:2004ct}. It is valid at long distances $r\ge 1/M_0\equiv r_0$, where $M_0$ is a scale corresponding to 
$\chi=1$. At this scale our effective theory is to be matched onto the pQCD; indeed at $\chi = 1$ our Lagrangian \eq{LAGR} becomes simply the Lagrangian of gluodynamics. The equation of motion of the dilaton field is
\beq\label{eqmot}
\frac{\vac}{m^2}\,\partial_\mu\left(
  e^{\chi/2}\,\partial_\mu\chi\right)
\,-\,
\frac{\vac}{4m^2}\,e^{\chi/2}\left(\partial_\mu\chi\right)^2\, +\,
\chi\, e^\chi\,\vac\,
+\,
\chi\, e^\chi\,c\,\frac{1}{4}\,(F_{\mu\nu}^a)^2\,
=\,0.
\eeq

The trace of energy-momentum  tensor can be  calculated directly from  \eq{LAGR} using  
\beq\label{trace}
\theta_\mu^\mu\,=\,g^{\mu\nu}\,\left(
 2\,\frac{\partial\lag}{\partial g^{\mu\nu}}\,-\,
g^{\mu\nu}\,\lag\right)
\,+\,
\frac{8\vac}{m^2}\,\partial_\mu^2\, e^{\chi/2},
\eeq
where the last term on the right hand side is the total
derivative. Using equation of motion of the dilaton field \eq{eqmot}
one arrives at 
\beq\label{sp}
\theta_\mu^\mu\,=\,-\,4\,\vac\, e^\chi\,-\,\chi\, e^\chi\,c\,(F_{\mu\nu}^a)^2\,.
\eeq
Since the vacuum at large distances corresponds to $\chi=0$, the color fluctuations represented by the second term in the r.h.s.\ of \eq{sp} decouple  -- the properties of the physical vacuum are determined by the fluctuations of the dilaton field. 

At small dilaton momenta its kinetic term is much smaller than the rest of terms in \eq{LAGR}. It catches up only at distances  $r\sim 1/m$. Therefore, if $r_0\gg 1/m$  the kinetic term remains small in the entire region of validity of the effective theory $r\ge r_0$.  We will verify later that this is indeed a valid assumption. Thus, we drope the kinetic term   in \eq{LAGR} and get
\begin{subequations}
\bea
\lag &\approx& \left(\vac\,+\,c\,\frac{1}{4} (F_{\mu\nu}^a)^2\right) e^\chi\,(1-\chi)\, -\,\frac{1}{4}\,(F^a_{\mu\nu})^2\,\\
&=& \vac \,e^\chi\,(1-\chi)- \frac{1}{4}\left[-e^\chi\,(1-\chi)\,c\, +1\right]\,(F^a_{\mu\nu})^2\label{01};
\eea
\end{subequations}
this action implies  that equation of motion for the dilaton is a constraint on the gluon field.
Extremizing the Lagrangian \eq{01} with respect to $\chi$ yields 
\beq\label{min.cond}
\chi\, e^\chi\,\left(\vac +c\,\frac{1}{4}\,(F^a_{\mu\nu})^2\right)|_\mathrm{min}\,=\,0\,.
\eeq
One solution to \eq{min.cond} is the physical vacuum at $\chi = 0$. However, once the chromo-electric gluon field $ (F^a_{\mu\nu})^2 < 0$ becomes sufficiently strong, \eq{min.cond} possesses a solution for any $\chi\neq 0$ if 
\beq\label{vac.min}
(F^a_{\mu\nu})^2|_\mathrm{min}=2(\vec B^{a 2}-\vec E^{a 2})|_\mathrm{min}= -4\vac\,c^{-1}\,.
\eeq

Assuming that the color field is created by static sources and neglecting the chromo-magnetic component of the field, the magnitude of the chromo-electric field from \eq{vac.min} is given by
\beq\label{chrom.el}
(\vec E^{a}_\mathrm{vac})^2 = \frac{2\vac}{c}.
\eeq
Unlike in pQCD, where $(F^a_{\mu\nu})^2|_\mathrm{min}=0$, the minimum of the effective theory corresponds to the finite chromoelectric field $E^a$.  
If the color field is weaker than \eq{chrom.el}, it is expelled from the vacuum, the ground state is at $\chi=0$, and the dynamics 
is described by interacting excitations above this vacuum -- the scalar fields $\chi$.

\medskip

Since the scalar field $\chi$ coupled to the chromo-electric field in the effective action \eq{LAGR} is color-singlet, the color dynamics at large distances becomes frozen. We thus will employ  
the quasi-Abelian gauge \cite{'tHooft:1981ht} $A_\mu^a=\phi(r) \,\delta^{a1}\,\delta^{\mu0}$,  with $r^2=x_ix^i$, $i=1,2,3$. In this gauge \eq{vac.min} reads 
\beq\label{eto.vac}
(\vec\nabla \phi_c(r))^2\,=\,2\vac \,c^{-1}\,\equiv\, \vec E_\mathrm{vac}^2
\eeq
with the solution $\phi_c(r)= |\vec E_\mathrm{vac}|\,r$ describing the classical background color field at large distances. 

Consider now the Coulomb potential induced in the presence of this background. We can define a 
renormalized at large distances potential by subtracting the background potential $\phi_c(r)$. 
This procedure corresponds to the \emph{infrared} renormalization.   
The Laplace equation for the renormalized Coulomb potential reads
\beq\label{laplace}
\vec\nabla ^2[\phi^\mathrm{ren}(r) - \phi_c(r)]\,=\, g\,\delta(\vec r)\,.
\eeq
It is solved by the Cornell potential 
\beq\label{renorm.pot}
\phi^\mathrm{ren}(r)\,=\, -\frac{g}{4\pi r}\,+\, |\vec E_\mathrm{vac}|\,r\,,
\eeq
where we used the identity $ \vec\nabla^2 r^{-1}=-4\pi \delta(\vec r) $. 
We have found that the effective theory \eq{LAGR} is confining.

\medskip

Let us emphasize again that Eq.~\eq{vac.min} requires the dilaton mass to vanish when the color flux tube is formed.   At large distances  $r\gg  r_0$ the quantum corrections build up to produce a significant dilaton mass which corresponds to the scalar glueball mass.  The appearance of the massless scalar particle -- the ``scalaron" -- is directly related to the condition \eq{vac.min} necessary for the confinement to occur. Therefore, the dilaton becoming a true massless Goldstone boson of broken scale symmetry signals the onset of confinement.

%%%%%%%%%%%%%%%%%%%%%%%%%%%%%%%%%%%%%%
\section{Matching to Perturbation Theory: The Effective Lagrangian at $\chi \simeq 1$.}\label{sec:el}

As we already discussed above, in the presence of strong chromo-electric fields exceeding \eq{chrom.el} the dilaton potential in \eq{LAGR} changes the sign, and the extremum at $\chi=0$ turns from a minimum to a maximum. The vacuum then shifts to the maximally allowed by 
our effective theory value $\chi = 1$. At $\chi = 1$ the effective Lagrangian \eq{LAGR} becomes the Lagrangian of gluodynamics. Since the effective dilaton potential $V(\chi)$ vanishes at $\chi = 1$, the energy density in this case is determined solely by the energy density of the gluon field, and has no contribution from the dilaton condensate. Likewise, in the expectation value of the trace of the energy-momentum tensor \eq{sp} at $\chi = 1$ the dilaton contribution and the contribution of the background flux field cancel each other. Therefore, to discuss the quantum dynamics at short distances corresponding to the vicinity of $\chi =1$ we have to expand the dilaton field around this value, and at the same time to expand the gluon field around the classical flux tube background fixed by \eq{chrom.el}.

To accomplish this, we introduce a new field $\phi(x)$ through 
\beq
\chi (x)= 1-v\,\phi(x)\,,\quad v\equiv\sqrt{\frac{m^2}{\vac\, e^{1/2}}}\,,
\eeq
and decompose the gluon field as $A_\mu^a\to \mathcal{A}^a_\mu+A_\mu^a$, where $A^a_\mu$ is a classical field satisfying \eq{eto.vac} and  $\mathcal{A}^a_\mu$ is a quantum fluctuation. The field strength decomposes as follows:
 \beq\label{LCHI1}
 F_{\mu\nu}^a\to F_{\mu\nu}^a + D_\mu \mathcal{A}_\nu^a-D_\nu \mathcal{A}_\mu^a+g^2f^{abc}\mathcal{A}_\mu^b\mathcal{A}_\nu^c\,,
\eeq
 where now $F_{\mu\nu}$ is the field strength of the classical field and $D_\mu=\partial_\mu-igA_\mu^at^a$ is the covariant derivative.  Expanding \eq{LAGR} in powers of $v\,\phi(x)$ to the second order we derive 
 \bea\label{LCHI11}
 \lag &=& \frac{1}{2}(\partial_\mu\phi)^2-\frac{1}{4}(F_{\mu\nu}^a + D_\mu \mathcal{A}_\nu^a-D_\nu \mathcal{A}_\mu^a+g^2f^{abc}\mathcal{A}_\mu^b\mathcal{A}_\nu^c)^2\nonumber\\
&&+e(v\,\phi-v^2\phi^2)\frac{1}{4}\bigg[2F_{\mu\nu}^a(D_\mu \mathcal{A}_\nu^a-D_\nu \mathcal{A}_\mu^a+g^2f^{abc}\mathcal{A}_\mu^b\mathcal{A}_\nu^c)\nonumber\\
&&
+(D_\mu \mathcal{A}_\nu^a-D_\nu \mathcal{A}_\mu^a+g^2f^{abc}\mathcal{A}_\mu^b\mathcal{A}_\nu^c)^2 \bigg] 
\eea
%%%%
\begin{figure}[ht] 
      \includegraphics[width=14cm]{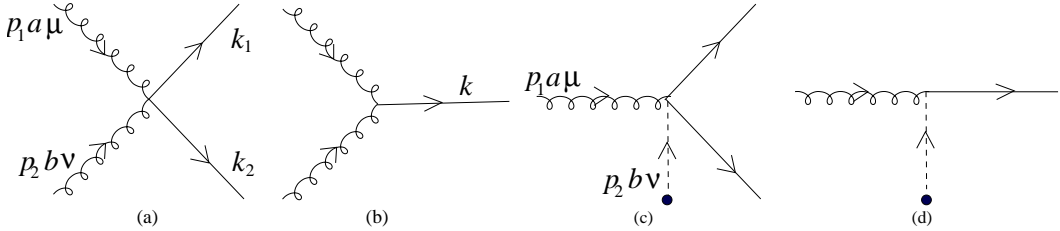} 
\caption{Interactions in \eq{LCHI11} to the order $g^0$. Helix line represents a gluon, dashed line -- classical color field, solid line -- dilaton.}
\label{fig:frules}
\end{figure}
%%%%%
Feynman diagrams to the order $g^0$ are displayed in \fig{fig:frules}. The corresponding rules read: 
\begin{subequations}
\bea
V_a&=& 2\,c\,e\,v^2\,[(p_1\cdot p_2)g_{\mu\nu}-p_{1\mu}p_{2\nu}]\,\delta^{ab}\,,\\
V_b&=& -\,c\,e\,v\,[(p_1\cdot p_2)g_{\mu\nu}-p_{1\mu}p_{2\nu}]\,\delta^{ab}\,,\\
V_c &=&  2\,c\,e\,v^2\,[A^a_\mu(p_2)(p_1\cdot p_2)-p_{2\mu}(p_1\cdot A^a(p_2))]\,,\\
V_d &=& -c\,e\,v\,[A^a_\mu(p_2)(p_1\cdot p_2)-p_{2\mu}(p_1\cdot A^a(p_2))]\,\,,
\eea
\end{subequations}
where $p$'s are gluon momenta as shown in \fig{fig:frules}. 
The effective Lagrangian \eq{LCHI11} can be used to compute non-perturbative corrections 
to QCD amplitudes.

\section{The gluon propagator}

For phenomenological applications it is convenient to determine the leading non-perturbative correction to the gluon propagator. 
Note that  the renormalized potential \eq{renorm.pot} satisfies the following equation
\beq
\vec\nabla^4\phi^\mathrm{ren}(\vec r)\,+\, 8\pi\,|\vec E_\mathrm{vac}| \,\delta(\vec r)\,=\, g\,\vec\nabla^2\,\delta(\vec r) ,
\eeq
which we Fourier transform into 
\beq
\phi^\mathrm{ren}(\vec k)\,=\, -\frac{g}{\vec k^2}\,-\,\frac{8\pi\,|\vec E_\mathrm{vac}| }{\vec k^4}\,.
\eeq
Therefore, the required expression for the renormalized gluon propagator in the Coulomb gauge is 
\beq\label{prop}
D(\vec k^2) =\frac{1}{k^4}\left(-\vec k^2-\mu^2\right)
\eeq
%\beq
%D(\vec k^2) = \frac{1}{\vec k^2}\,\frac{1}{1-\mu^2/ \vec k^2}\,,\quad \vec k^2\gg \mu^2\,.
%\eeq
where we denoted 
\beq\label{mass}
\mu^2=\frac{8\pi\,|\vec E_\mathrm{vac}|}{g}\,.
\eeq
The gluon propagator with a $\sim 1/\vec k^4$ behavior in the infrared has been advocated before  \cite{Gribov:1977wm,Zwanziger:2002sh}. In particular, it has been argued \cite{Shirkov:1997wi,Chetyrkin:1998yr} that such a behavior provides a way of extending the reach of perturbative QCD into the semi-hard region.

%%%%%%%%%%%%%%%%%%
\section{Strong coupling in the infrared}\label{sec:sc}

We wish now to examine the behavior of the strong coupling at long distances. Let us concentrate on the color sector of the effective Lagrangian \eq{LAGR} and  introduce the electric susceptibility of QCD vacuum
\beq\label{eps}
\epsilon(\chi)\equiv Z(\chi)= 1-c\,e^\chi\, (1-\chi)\,,
\eeq
If we now rescale the gluon field potentials as $\bar A^a_\mu = gA^a_\mu$ and $\bar F^2 = g^2F^2$, the Lagrangian \eq{LAGR} can be written as 
\beq\label{resc}
\lag \,=\, \frac{\vac}{m^2}\,\frac{1}{2}\,e^{\chi/2}\, (\partial_{\mu}\chi)^2\,
+\,\vac\,e^\chi\,(1-\chi)\,-\frac{1}{4}\,\frac{\epsilon(\chi)}{g^2}\,(\bar F^a_{\mu\nu})^2\,,
\eeq
The form of \eq{resc} implies the behavior of the renormalized strong coupling in the presence of dilaton background:
\beq\label{run}
\as(\chi(r))=\frac{\as(M_0)}{\epsilon(\chi)}=\frac{\as(M_0)}{1-\, c\,e^\chi\, (1-\chi)}\,.
\eeq
At $\chi =1$ the coupling is $\as(M_0)$, which is consistent with our procedure of matching onto perturbation theory at scale $M_0$ (and the corresponding distance $r_0 = M_0^{-1}$). 
Note that in the physical vacuum $\chi=0$, coupling constant becomes 
\beq\label{run.infty}
\as(\chi\to 0)=\frac{\as(M_0)}{1-c}\ ;
\eeq 
since $c \ll 1$, the effective coupling in our effective theory remains quite small at large distances, and the leading quantum corrections arise from the interactions with the dilaton fields.  

\medskip

To determine the distance $r_0$ at which matching to the pQCD takes place, we write down the Yang-Mills equations in the presence of the point-like source\cite{Pagels:1978dd}:
\beq\label{YM}
D_\nu^{ab}\left(\epsilon(\chi)\, \bar F_{\nu\mu}^b\right)\,=\, j_\mu^a\,,\quad
D_\mu^{ab}=\delta^{ab}\partial_\mu+f^{bac}\bar A_\mu^c\,,
\eeq
where $\bar A^a_\mu = gA^a_\mu$ and $\bar F^2 = g^2F^2$. In the quasi-Abelian gauge we get for the radial component of the electric field of a point source
\beq\label{mod.coul}
\bar E=\frac{1}{4\pi r^2\, \epsilon(\chi)} 
\eeq
To find $r_0$ where our effective theory matches onto the pQCD we note that the classical minimum corresponds to $E=E_\mathrm{vac}$ (see \eq{eto.vac}) 
and set $\chi=1$, that is $\epsilon(\chi)=1$. We obtain
 \beq\label{r0det}
 r_0=\sqrt{\frac{g}{4\pi  E_\mathrm{vac}}}\,.
 \eeq
 
\section{The structure of the flux tube}
 
 We can now determine  the profile of the 
 energy density stored in the gluon--dilaton configuration as a function of coordinate $r$. For the energy density we obtain from the Lagrangian \eq{LAGR}:
\begin{eqnarray}
\theta^{00}(x)&=&
\frac{\vac}{2m^2}\,[(\partial_0\chi)^2+(\partial_i\chi)^2]\,e^{\chi/2}
-\,\vac\,e^\chi\,(1\,-\,\chi)\,
\nonumber\\
&& 
+\,\left(-F^{a0\lambda}F^{a0}_{\quad\lambda}\,+
\,\frac{1}{4}\,\,
(F_{\lambda\sigma}^a)^2\right)\,\left(
1\,-\,c\, e^\chi\,(1\,-\,\chi)\right),
\end{eqnarray}
where $i=1,2,3$. For the constant dilaton field and the Coulomb gauge  we work in we have
\beq\label{w1}
\theta^{00}=-\vac\,e^\chi\,(1\,-\,\chi)+\frac{1}{2}\bar E^2\,\frac{1}{g^2}\,\left( 1-c\, e^\chi\,(1-\chi)\right)\,.
\eeq
Setting $E=E_\mathrm{vac}$ in \eq{mod.coul} we derive near $r=r_0$ (where the Coulomb law \eq{mod.coul}, rather than the string potential, holds)
\beq\label{w2}
\theta^{00}(r)=-\vac\, c^{-1}+ 2\,\vac \,\frac{1}{4\pi r^2 \bar E_0^2}\,c^{-1}=
-\vac\left(1 - \frac{3r_0^2}{ r^2 }\right)\,, \quad r\approx r_0\,.
\eeq
After the subtraction of the vacuum term in \eq{w2} the energy density decreases as $~1/r^2$ implying that the total energy stored in the infrared gluon configuration is linear in distance. This is yet another way to see the formation of the flux tube at (rather short) distances $r\approx r_0$. We see also that the energy density changes sign at  $r=\sqrt{3}r_0$ which can be identified with the radius of the flux tube, while $\vac$ is the difference in the energy density inside and outside of the tube, analogous to the bag constant. 

 %%%%%%%
\section{Relation to 't~Hooft's ``perturbative confinement"}  \label{sec:thooft}

Lagrangian \eq{01} is a particular  realization of 't~Hooft's ``perturbative confinement" \cite{'t Hooft:2002wz} approach.   It is instructive to review the arguments given in \cite{'t Hooft:2002wz}. The most general renormalized Lagrangian $\mathcal{L}=f(F_{\mu\nu}^2)$ can be written in the form
\beq\label{h.lagr}
\mathcal{L}(A,\phi)=-\frac{1}{4}\,Z(\phi)\, F_{\mu\nu}^2\, -\,V(\phi)+J_\mu A_\mu\,,
\eeq
where $\phi$ is a scalar field with self-action $V(\phi)$ and $J_\mu$ is an external source. We assume the quasi-Abelian gauge thus dropping the color indices.  The requirement of confinement imposes a restriction on functions $V(\phi)$ and $Z(\phi)$. To derive these conditions we introduce the electric displacement field $\vec D$ 
\beq
\vec D=Z(\phi)\vec E\,.
\eeq
Then extremizing \eq{h.lagr} with respect to $\phi$ yields
\beq\label{h.extr}
\frac{1}{2}D^2=-\frac{\partial V}{\partial (1/Z)}
\eeq

Consider now the energy density $U(D,\phi)$ stored in a particular configuration of $D$ and $\phi$ fields
\beq\label{h.energy}
U=\vec D\cdot \partial_0 \vec A-\mathcal{L}=\frac{1}{2}\frac{D^2}{Z(\phi)}+V(\phi)+J_0A_0\,.
\eeq
Variation in a configuration results in change
\beq\label{h.dU}
dU=\frac{1}{2}D^2\,d\frac{1}{Z}\,+\, \frac{1}{Z}\,d\left(\frac{1}{2}D^2\right)\,+
\, dV =\frac{1}{Z}\,d\left(\frac{1}{2}D^2\right)\,.
\eeq

Now, confinement occurs when the most energetically favorable configuration corresponds to the linearly rising potential
\beq\label{h.conf}
U(\vec D)=\rho\, |\vec D|
\eeq
at least at small values  of $|\vec D|$, i.e. in the transition region between perturbation theory and confinement. In  \eq{h.conf} $\rho$ can be readily identified as a string tension. In this region where \eq{h.conf} holds we have using \eq{h.dU}
\beq\label{h.z}
\frac{1}{Z}=\frac{dU}{d(D^2/2)}\approx \frac{d\rho}{d(D^2/2)}=\frac{\rho}{D}\,.
\eeq
With the help of \eq{h.extr} we derive
\beq
V=-\int \,\frac{1}{2}\, D^2\, d(1/Z)\,\approx\, \frac{1}{2}\rho\, D=\frac{1}{2}\rho^2 Z\,.
\eeq

Let us now rewrite \eq{01} in a suggestive form
\beq
\lag 
= -V(\chi) - Z(\chi) \frac{1}{4}\,(F^a_{\mu\nu})^2\,.
\eeq
were the following notations were introduced 
 \beq
 Z(\chi)=-e^\chi\,(1-\chi) \,c\,+1\,,\quad V(\chi)=-\vac \,e^\chi\,(1-\chi)\,.
 \eeq
%%%%
\begin{figure}[ht] 
    \includegraphics[width=10cm]{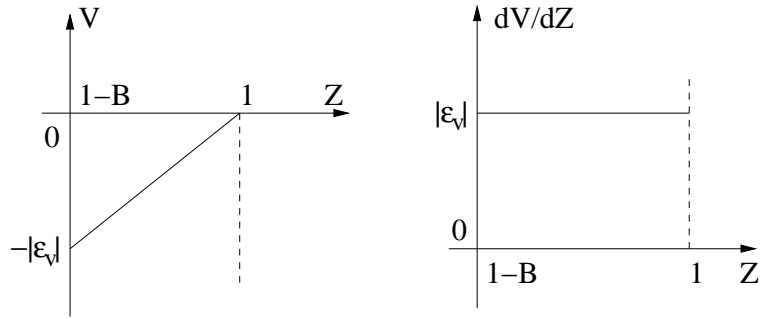} 
\caption{A particular realization of the 't Hooft's perturbative confinement \cite{'t Hooft:2002wz} as explained in the text.}
\label{fig:vz}
\end{figure}
%%%%%
 The resulting values of $V$ and $\partial{V}/{\partial Z}$ are displayed in \fig{fig:vz} for  $\frac{\beta(g)}{2g}=1$. 
 
At large values $D$ the energy density must take the perturbative form $U=D^2/2$. Therefore,   \eq{h.z} it implies that  $Z\to 1$ which in turn corresponds to $\chi=1$. 
  
\section{Numerical estimates}

Let us check whether the relations following from our effective theory make sense phenomenologically. To do this, we have to choose numerical values for the parameters entering the Lagrangian \eq{LAGR}: the dilaton mass $m$ and the vacuum energy density $\vac$. In addition, since the effective theory \eq{LAGR} is non-renormaliable, we also have to specify the value of the matching scale $M_0$. This value will also determine the value of the strong coupling $\as(M_0)$ thus defining the constant $c$. Since we have not introduced light quarks so far, our estimates will be applicable to pure gluodynamics. 

In gluodynamics, the dilaton has to be identified with the scalar glueball;  lattice QCD gives the mass of $m = 1.5 \div 1.7$ GeV, and we pick the value $m = 1.6$ GeV. 
In accord with our previous work we choose $M_0 = 2 \ {\rm GeV}$ for  the scale $M_0$ at which the perturbation theory starts to apply.  The corresponding value of the QCD coupling is $\as(M_0) \simeq 0.35$, and $c=|\beta(g_0)|/2g_0 \simeq b\ \as(M_0) / ( 8 \pi) \simeq 0.15$; we used $b=11$ as appropriate for gluodynamics with $N_c = 3$. 

The value of the vacuum energy density is somewhat uncertain; in gluodynamics it is related to the gluon condensate by the relation $\vac = 11/32 \langle (\as/\pi) F^2 \rangle$. The original value of the gluon condensate is $\langle (\as/\pi) F^2 \rangle = 0.012\ {\rm GeV}^4$ \cite{Shifman:1978by}. The latest analysis \cite{Ioffe:2005ym} (including in particular an updated knowledge of $\as$) yields a significantly smaller value 
$\langle (\as/\pi) F^2 \rangle = 0.005\pm0.004\ {\rm GeV}^4$ leading to $\vac = 0.0017\pm0.0014$. In this situation we will work backwards and pick the 
value of $\vac$ using the relation \eq{chrom.el}. Eq. \eq{chrom.el} relates the  value of the vacuum chromo-electric field $E_\mathrm{vac}$ (which according to \eq{renorm.pot} plays the role of string tension in our approach) to the values of $\vac$ and $c$. Choosing $E_\mathrm{vac} \simeq 800 \ {\rm MeV/fm}$  for the string tension and  the value $c = 0.15$ inferred above we get for the vacuum energy density from \eq{chrom.el} 
\beq
\vac \simeq 0.0019 \ {\rm GeV}^4,
\eeq
a value consistent with the analysis \cite{Ioffe:2005ym}. The radius of the flux tube can now be found from 
\eq{w2} and  \eq{r0det}; numerically, we find $r_\mathrm{tube} \simeq 0.2$ fm, a reasonable value.

Admittedly there is a significant uncertainty in the values of the parameters $\vac$ and $M_0$ of the effective Lagrangian \eq{LAGR}. However reasonable choices of these parameters yield  phenomenologically sound values of the string tension, the radius of the flux tube and the value of the effective coupling in the infrared.

%%%%%%%%%%%  %%%%%%%%%%%%%%

\section{Discussion}\label{sec:disc}
 
An effective theory of broken scale symmetry of QCD given by the Lagrangian \eq{LAGR} which we have advocated here and elsewhere \cite{klt,Kharzeev:2004ct} possesses a number of remarkable properties: it yields confinement, and links the formation of color tube to the emergence of a massless scalar excitation -- ``the scalaron". 
Let us describe the relevant degrees of freedom in this theory at different distances in simple physical terms. 
At large distances, the gluons are bound into massive scalar glueballs. At shorter distances 
(inside the hadron) the quasi-Abelian color flux tube is found, and the dominant degree of freedom are 
the massless dilaton excitations, the scalarons. At still shorter distances we encounter gluons with non-perturbative interactions induced by the coupling to scalarons. Finally, at very short distances we match onto the usual QCD perturbation theory. 

The basic idea of modifying the dynamics of gluon fields at large distances is of course not new, and several approaches of that kind were mentioned in the Introduction. The distinctive feature of the approach advocated in this paper is that our effective Lagrangian is fixed entirely by scale anomaly. Knowing the structure of the effective Lagrangian allowed us to establish that it possesses the property of confinement, and to link the formation of color confining flux tube to the emergence of massless scalar excitation. 

In our opinion the effective theory \eq{LAGR} provides a way for systematically computing non-perturbative corrections to various amplitudes; we think it is worthwhile to pursue the phenomenological applications both at zero and finite temperature. We have already discussed the IR behavior of the strong coupling in this framework \cite{Kharzeev:2004ct} and the effect of soft gluon emission on the structure of the leading Regge singularity at high energies, i.e. the Pomeron \cite{klt}. This approach has an interesting implication also for the spin structure of the nucleon -- since the gluons at semi-hard scales are bound in this picture into scalar (spin-singlet) scalaron states, there should be no contribution from gluons to the spin of the hadron at semi-hard scales 
where the perturbative evolution is initiated. This is perhaps consistent with the preliminary results 
on the fraction of the proton's spin carried by gluons from RHIC and elsewhere.

 It has already been demonstrated that the dilaton excitations near the critical temperature are responsible for the anomalous bulk viscosity \cite{Kharzeev:2007wb}. Likewise it determines the behavior of the trace of energy-momentum tensor at temperatures above $T_c$ and may give an important contribution to the parton energy loss. In the deconfined phase color charges are screened at distances of the order of the inverse Debye mass, while the massless dilatons mediate the long-range strong force with possible important influence on the global dynamics of the quark-gluon plasma.
 We leave the development of these applications for the future.

%%%%%%%%%%%%%%%%%%%%%%%%%
\acknowledgments
The work of D.K. was supported by the U.S. Department of Energy under Contract No. DE-AC02-98CH10886. 
 K.T.  is supported in part by the U.S. Department of Energy Grant No. DE-FG02-87ER40371; he would also like to thank
RIKEN, BNL and the U.S. Department of Energy (Contract
No. DE-AC02-98CH10886) for providing the facilities essential for the
completion of this work.  This research of E.L.  was supported 
in part by the Israel Science Foundation, founded by the Israeli Academy of Science 
and Humanities and  by BSF grant $\#$ 20004019.

%%%%%%%%%%%%%%%%%%%%%%%%%%%%%%%%%%%%%%
%%%%%%%%%%%%%%%%%%%%%%%%%%%%%%%%%%%%%


\begin{thebibliography}{70}

%\cite{Gribov:1977wm}
\bibitem{Gribov:1977wm}
  V.~N.~Gribov,
  %``Quantization of non-Abelian gauge theories,''
  Nucl.\ Phys.\  B {\bf 139}, 1 (1978).
  %%CITATION = NUPHA,B139,1;%%

\bibitem{scale1}
  J.~R.~Ellis,
  %``Aspects of conformal symmetry and chirality,''
  Nucl.\ Phys.\  B {\bf 22}, 478 (1970);\\
  %%CITATION = NUPHA,B22,478;%%
  R.~J.~Crewther,
  %``BROKEN SCALE INVARIANCE IN THE WIDTH OF A SINGLE DILATON,''
  Phys.\ Lett.\  B {\bf 33}, 305 (1970); \\
   %%CITATION = PHLTA,B33,305;%%
   M.~S.~Chanowitz and J.~R.~Ellis,
  %``Canonical Anomalies And Broken Scale Invariance,''
  Phys.\ Lett.\  B {\bf 40}, 397 (1972);\\
  %%CITATION = PHLTA,B40,397;%%
  %\cite{Schechter:1980ak}
  J.~Schechter,
  %``Effective Lagrangian With Two Color Singlet Gluon Fields,''
  Phys.\ Rev.\  D {\bf 21}, 3393 (1980).
  %%CITATION = PHRVA,D21,3393;%%
  
\bibitem{scale2}
J.~C.~Collins, A.~Duncan and S.~D.~Joglekar,
  %``Trace And Dilatation Anomalies In Gauge Theories,''
  Phys.\ Rev.\  D {\bf 16}, 438 (1977);\\
  %%CITATION = PHRVA,D16,438;%%  
 %\cite{Nielsen:1977sy}
  N.~K.~Nielsen,
  %``The Energy Momentum Tensor In A Nonabelian Quark Gluon Theory,''
  Nucl.\ Phys.\  B {\bf 120}, 212 (1977).
  %%CITATION = NUPHA,B120,212;%%


%\cite{Gross:1973id}
\bibitem{Gross:1973id}
  D.~J.~Gross and F.~Wilczek,
  %``ULTRAVIOLET BEHAVIOR OF NON-ABELIAN GAUGE THEORIES,''
  Phys.\ Rev.\ Lett.\  {\bf 30}, 1343 (1973);\\
  %%CITATION = PRLTA,30,1343;%%
  %\cite{Politzer:1973fx}
%\bibitem{Politzer:1973fx}
  H.~D.~Politzer,
  %``RELIABLE PERTURBATIVE RESULTS FOR STRONG INTERACTIONS?,''
  Phys.\ Rev.\ Lett.\  {\bf 30}, 1346 (1973).
  %%CITATION = PRLTA,30,1346;%%
  
\bibitem{schechter}
J.~Schechter,
%``Effective Lagrangian With Two Color Singlet Gluon Fields,''
Phys.\ Rev.\ D {\bf 21}, 3393 (1980);\\
%%CITATION = PHRVA,D21,3393;%%
A.~Salomone, J.~Schechter and T.~Tudron,
%``Properties Of Scalar Gluonium,''
Phys.\ Rev.\ D {\bf 23}, 1143 (1981);\\
%%CITATION = PHRVA,D23,1143;%%
H.~Gomm, P.~Jain, R.~Johnson and J.~Schechter,
%``Bag Formation In A Chiral Model,''
Phys.\ Rev.\ D {\bf 33}, 801 (1986).
%%CITATION = PHRVA,D33,3476;%%

\bibitem{Migdal:1982jp}
  A.~A.~Migdal and M.~A.~Shifman,
  %``Dilaton Effective Lagrangian In Gluodynamics,''
  Phys.\ Lett.\  B {\bf 114}, 445 (1982).
  %%CITATION = PHLTA,B114,445;%%

%\cite{Chen:2005mg}
\bibitem{Chen:2005mg}
  Y.~Chen {\it et al.},
  %``Glueball spectrum and matrix elements on anisotropic lattices,''
  Phys.\ Rev.\  D {\bf 73}, 014516 (2006)
  [arXiv:hep-lat/0510074].
  %%CITATION = PHRVA,D73,014516;%%


\bibitem{klt}
D.~Kharzeev, E.~Levin and K.~Tuchin,
%``Classical gluodynamics in curved space-time and the soft pomeron,''
Phys.\ Lett.\ B {\bf 547}, 21 (2002)
[arXiv:hep-ph/0204274].
%%CITATION = HEP-PH 0204274;%%


%\cite{Kharzeev:2004ct}
\bibitem{Kharzeev:2004ct}
  D.~Kharzeev, E.~Levin and K.~Tuchin,
  %``QCD in curved space-time: A conformal bag model,''
  Phys.\ Rev.\  D {\bf 70}, 054005 (2004)
  [arXiv:hep-ph/0403152].
  %%CITATION = PHRVA,D70,054005;%%

%\cite{Zwanziger:2002sh}
\bibitem{Zwanziger:2002sh}
  D.~Zwanziger,
  %``No confinement without Coulomb confinement,''
  Phys.\ Rev.\ Lett.\  {\bf 90}, 102001 (2003)
  [arXiv:hep-lat/0209105].
  %%CITATION = PRLTA,90,102001;%%

%\cite{Dokshitzer:2004ie}
\bibitem{Dokshitzer:2004ie}
  Y.~L.~Dokshitzer and D.~E.~Kharzeev,
  %``The Gribov conception of quantum chromodynamics,''
  Ann.\ Rev.\ Nucl.\ Part.\ Sci.\  {\bf 54}, 487 (2004)
  [arXiv:hep-ph/0404216].
  %%CITATION = ARNUA,54,487;%%

%\cite{Gaete:2007zn}
\bibitem{Gaete:2007zn}
  P.~Gaete and E.~Spallucci,
  %``Confinement from gluodynamics in curved space-time,''
  Phys.\ Rev.\  D {\bf 77}, 027702 (2008)
  [arXiv:0707.2738 [hep-th]].
  %%CITATION = PHRVA,D77,027702;%%

%\cite{Kharzeev:2005iz}
\bibitem{Kharzeev:2005iz}
  D.~Kharzeev and K.~Tuchin,
  %``From color glass condensate to quark gluon plasma through the event
  %horizon,''
  Nucl.\ Phys.\  A {\bf 753}, 316 (2005)
  [arXiv:hep-ph/0501234];\\
  %%CITATION = NUPHA,A753,316;%%
  D.~Kharzeev, E.~Levin and K.~Tuchin,
  %``Multi-particle production and thermalization in high-energy QCD,''
  Phys.\ Rev.\  C {\bf 75}, 044903 (2007)
  [arXiv:hep-ph/0602063].
  %%CITATION = PHRVA,C75,044903;%%
  
 %\cite{Castorina:2007eb}
\bibitem{Castorina:2007eb}
  P.~Castorina, D.~Kharzeev and H.~Satz,
  %``Thermal Hadronization and Hawking-Unruh Radiation in QCD,''
  Eur.\ Phys.\ J.\  C {\bf 52}, 187 (2007)
  [arXiv:0704.1426 [hep-ph]].
  %%CITATION = EPHJA,C52,187;%% 
  
%\cite{Chodos:1974je}
\bibitem{Chodos:1974je}
  A.~Chodos, R.~L.~Jaffe, K.~Johnson, C.~B.~Thorn and V.~F.~Weisskopf,
  %``A New Extended Model Of Hadrons,''
  Phys.\ Rev.\  D {\bf 9}, 3471 (1974).
  %%CITATION = PHRVA,D9,3471;%%

%\cite{Friedberg:1976eg}
\bibitem{Friedberg:1976eg}
  R.~Friedberg and T.~D.~Lee,
  %``Fermion Field Nontopological Solitons. 1,''
  Phys.\ Rev.\  D {\bf 15}, 1694 (1977);
  %%CITATION = PHRVA,D15,1694;%%
%``Fermion Field Nontopological Solitons. 2. Models For Hadrons,''
  Phys.\ Rev.\  D {\bf 16}, 1096 (1977).
  %%CITATION = PHRVA,D16,1096;%%
 
 
 %\cite{Kogut:1974sn}
\bibitem{Kogut:1974sn}
  J.~B.~Kogut and L.~Susskind,
  %``Vacuum Polarization And The Absence Of Free Quarks In Four-Dimensions,''
  Phys.\ Rev.\  D {\bf 9}, 3501 (1974).
  %%CITATION = PHRVA,D9,3501;%%
 
 %\cite{Pirner:1991im}
\bibitem{Pirner:1991im}
  H.~J.~Pirner,
  %``The Color dielectric model of QCD,''
  Prog.\ Part.\ Nucl.\ Phys.\  {\bf 29}, 33 (1992).
  %%CITATION = PPNPD,29,33;%%

%\cite{Dick:1997ju}
\bibitem{Dick:1997ju}
  R.~Dick,
  %``The Coulomb potential in gauge theory with a dilaton,''
  Phys.\ Lett.\  B {\bf 397}, 193 (1997)
  [arXiv:hep-th/9701047];
  %%CITATION = PHLTA,B397,193;%%
  %``Vector and scalar confinement in gauge theory with a dilaton,''
  Phys.\ Lett.\  B {\bf 409}, 321 (1997)
  [arXiv:hep-ph/9706278];  
  %``Confinement from a massive scalar in {QCD},''
  Eur.\ Phys.\ J.\  C {\bf 6}, 701 (1999)
  [arXiv:hep-ph/9803209];\\
  A.~Wereszczynski,
  %``Universality of the linear potential in effective models for the low energy
  %QCD coupled with the dilaton field,''
  Phys.\ Lett.\  B {\bf 570}, 260 (2003)
  [arXiv:hep-ph/0310236];\\
  A.~Wereszczynski and M.~Slusarczyk,
  %``Non-Abelian color dielectric: Towards the effective model of the low
  %energy QCD,''
  Eur.\ Phys.\ J.\  C {\bf 39}, 185 (2005)
  [arXiv:hep-ph/0405148];\\
M.~Chabab,
  %``On the implications of a dilaton in gauge theory,''
  Int.\ J.\ Mod.\ Phys.\  A {\bf 22}, 5717 (2008)
  [arXiv:0709.1226 [hep-ph]].
  %%CITATION = IMPAE,A22,5717;%%

%\cite{'t Hooft:2002wz}
\bibitem{'t Hooft:2002wz}
  G.~'t Hooft,
  %``Perturbative confinement,''
  Nucl.\ Phys.\ Proc.\ Suppl.\  {\bf 121}, 333 (2003)
  [arXiv:hep-th/0207179].
  %%CITATION = NUPHZ,121,333;%%

%\cite{Kharzeev:2007wb}
\bibitem{Kharzeev:2007wb}
  D.~Kharzeev and K.~Tuchin,
  %``Bulk viscosity of QCD matter near the critical temperature,''
  arXiv:0705.4280 [hep-ph].
  %%CITATION = ARXIV:0705.4280;%%

%\cite{Karsch:2007jc}
\bibitem{Karsch:2007jc}
  F.~Karsch, D.~Kharzeev and K.~Tuchin,
  %``Universal properties of bulk viscosity near the QCD phase transition,''
  Phys.\ Lett.\  B {\bf 663}, 217 (2008)
  [arXiv:0711.0914 [hep-ph]].
  %%CITATION = PHLTA,B663,217;%%

%\cite{Meyer:2007dy}
\bibitem{Meyer:2007dy}
  H.~B.~Meyer,
  %``A calculation of the bulk viscosity in SU(3) gluodynamics,''
  Phys.\ Rev.\ Lett.\  {\bf 100}, 162001 (2008)
  [arXiv:0710.3717 [hep-lat]].
  %%CITATION = PRLTA,100,162001;%%

%\cite{Novikov:1981xj}
\bibitem{Novikov:1981xj}
  V.~A.~Novikov, M.~A.~Shifman, A.~I.~Vainshtein and V.~I.~Zakharov,
  %``Are All Hadrons Alike?,''
  Nucl.\ Phys.\  B {\bf 191}, 301 (1981).
  %%CITATION = NUPHA,B191,301;%%
%\cite{Migdal:1982jp}

%\cite{Shifman:1978by}
\bibitem{Shifman:1978by}
  M.~A.~Shifman, A.~I.~Vainshtein and V.~I.~Zakharov,
  %``QCD And Resonance Physics: Applications,''
  Nucl.\ Phys.\  B {\bf 147}, 448 (1979).
  %%CITATION = NUPHA,B147,448;%%

%\cite{'tHooft:1981ht}
\bibitem{'tHooft:1981ht}
 G.~'t Hooft,
 %``Topology Of The Gauge Condition And New Confinement Phases In Nonabelian
 %Gauge Theories,''
 Nucl.\ Phys.\  B {\bf 190}, 455 (1981).
 %%CITATION = NUPHA,B190,455;%%

%\cite{Pagels:1978dd}
\bibitem{Pagels:1978dd}
  H.~Pagels and E.~Tomboulis,
  %``Vacuum Of The Quantum Yang-Mills Theory And Magnetostatics,''
  Nucl.\ Phys.\  B {\bf 143}, 485 (1978).
  %%CITATION = NUPHA,B143,485;%%

%\cite{Shirkov:1997wi}
\bibitem{Shirkov:1997wi}
  D.~V.~Shirkov and I.~L.~Solovtsov,
  %``Analytic model for the QCD running coupling with universal  alpha(s)-bar(0)
  %value,''
  Phys.\ Rev.\ Lett.\  {\bf 79}, 1209 (1997)
  [arXiv:hep-ph/9704333].
  %%CITATION = PRLTA,79,1209;%%
  
  
%\cite{Chetyrkin:1998yr}
\bibitem{Chetyrkin:1998yr}
  K.~G.~Chetyrkin, S.~Narison and V.~I.~Zakharov,
  %``Short-distance tachyonic gluon mass and 1/Q**2 corrections,''
  Nucl.\ Phys.\  B {\bf 550}, 353 (1999)
  [arXiv:hep-ph/9811275].
  %%CITATION = NUPHA,B550,353;%%


%\cite{Ioffe:2005ym}
\bibitem{Ioffe:2005ym}
  B.~L.~Ioffe,
  %``QCD at low energies,''
  Prog.\ Part.\ Nucl.\ Phys.\  {\bf 56}, 232 (2006)
  [arXiv:hep-ph/0502148].
  %%CITATION = PPNPD,56,232;%%
\end{thebibliography}
\end{document}